\documentclass[aps,amssymb,amsmath,twocolumn, pra, superscriptaddress]{revtex4}
\usepackage{graphicx}
\usepackage{epsfig}
\usepackage{dcolumn}
\usepackage{bm}
\usepackage{bbm}
\usepackage{hyperref}
\usepackage[up]{subfigure}

\newcommand{\be}{\begin{equation}}
\newcommand{\ee}{\end{equation}}
\newcommand{\bc}{\begin{center}}
\newcommand{\ec}{\end{center}}
\newcommand{\bi}{\begin{itemize}}
\newcommand{\ei}{\end{itemize}}
\newcommand{\bea}{\begin{eqnarray}}
\newcommand{\eea}{\end{eqnarray}}
\begin{document}
\title{Optimizing the  discrete time quantum walk using a $SU(2)$ coin}
\author{C. M. \surname{Chandrashekar}}
\affiliation{Institute for Quantum  Computing, University of Waterloo, ON, N2L 3G1, Canada}
\author{R. \surname{Srikanth}}
\affiliation{Poornaprajna Institute of Scientific Research, Devanahalli, Bangalore 562 110, India}
\affiliation{Raman Research Institute, Sadashiva  Nagar,  Bangalore,  India}
\author{Raymond \surname{Laflamme}}  
\affiliation{Institute for Quantum Computing, University  of Waterloo, ON,  N2L 3G1,  Canada} 
\affiliation{Perimeter Institute for Theoretical Physics, Waterloo, ON, N2J 2W9, Canada}
\vskip 4cm
\begin{abstract}
We present  a generalized version  of the discrete time  quantum walk,
using the $SU(2)$ operation as  the quantum coin.  By varying the coin
parameters,  the quantum walk  can be  optimized for  maximum variance
subject  to  the  functional  form  $\sigma^2 \approx  N^2$  and  the
probability distribution in the position  space can be biased. We also
discuss the variation in measurement entropy with the variation of the
parameters in the  $SU(2)$ coin.  Exploiting this we  show how quantum
walk can be  optimized for improving mixing time  in an $n$-cycle and for
quantum walk search.
\end{abstract}
\maketitle
\preprint{Version}
\section{Introduction}
The discrete time quantum walk has a very similar structure to that of
the classical random walk - a coin  flip and a subsequent shift - but the
behaviour is strikingly different because of quantum interference. The
variance $\sigma^{2}$ of the quantum  walk is known to grow quadratically with the number of steps $N$, $\sigma^{2}\propto  N^{2}$, compared  to the
linear growth, $\sigma^{2}\propto  N$, for  the classical random walk \cite{aharonov,kemp,ashwin, andris}. This has motivated  the  exploration  for a  new  and  improved  quantum  search algorithms,  which  under certain  conditions  are exponentially  fast compared to the  classical analog \cite{childs}. Environmental effects 
on the quantum walk \cite{chandra07} and the role of the quantum walk to speed
up the physical process, such as the quantum phase transition have been 
explored  \cite{chandra07a}. Experimental  implementation of  the  
quantum walk  has been  reported \cite{ryan}  and various other schemes 
for a physical realization have been proposed \cite{travaglione}.
\par
The quantum walk of a  particle initially in a symmetric superposition
state $|\Psi_{in}\rangle$ using  a single-variable  parameter $\theta$ in
the  unitary  operator,  $U_{\theta} \equiv \left( \begin{array}{clcr}
 \cos(\theta)  & &  \sin(\theta)   \\
\sin(\theta)  & &  -\cos(\theta)
\end{array} \right)$,   as  quantum  coin  returns  the
symmetric probability distribution in  the position space.  The change
in  the parameter $\theta$  is known  to affect  the variation  in the
variance,  $\sigma^{2}$  \cite{ashwin}.   It  has been  reported  that
obtaining a symmetric distribution depends  largely on the initial state
of the particle \cite{ashwin, andris, tregenna}.
\par
In this paper, the discrete  time quantum walk has  been generalized using  the $SU(2)$  operator with  three Caley-Klein  parameters $\xi$, $\theta$ and $\zeta$  as  the  quantum  coin. We  show  that  the  variance can be varied by changing the parameter $\theta$, $\sigma^{2} \approx (1-\sin(\theta))N^{2}$ and the parameters $\xi$ and $\zeta$  introduce asymmetry in the position-space probability  distribution even if the initial state of the  particle  is  in  symmetric superposition.
This asymmetry in the probability distribution is similar to the distribution obtained for a walk on a particle initially in a non-symmetric superposition state. We  discuss  the variation  of measurement  entropy in  position space  with  the three parameters.  Thus, we also show that the quantum walk can be optimized for  the  maximum  variance,  for  applications  in  search  algorithm, improving  mixing time  in  an  $n$-cycle or  general  graph and  other processes using a generalized $SU(2)$ quantum coin. The combination of the measurement entropy  and three parameters in the  $SU(2)$ coin can be  optimized  to  fit  the  physical  system  and  for  the  relevant applications of the quantum walk on general graphs. This paper discuss the effect of $SU(2)$ coin on quantum walk with particle initially in symmetric superposition state. The $SU(2)$ coin will have a similar influence on a particle starting with other initial states but with an additional decrease in the variance by a small amount.    
\par
The  paper is  organized  as  follows.  Section \ref{qrw}  introduces to the discrete time quantum (Hadamard) walk.  Section \ref{gen-qw} discusses the generalized version of the quantum  walk using the arbitrary three-parameter $SU(2)$ quantum coin. The effect of three parameters on the variance of the quantum walker is discussed, and the functional dependence of the variance due to parameter $\theta$ is shown. The variation of the entropy of the measurement in position space after implementing the quantum walk using different values of $\theta$ is discussed in Sec. \ref{entropy}. 
Section \ref{cycle} and \ref{search} discuss optimization of the mixing time of the quantum walker on the $n-$cycle and the search using a quantum walk. Section 
\ref{conclusion} concludes with a summary.
\section{Hadamard  walk}
\label{qrw}
\begin{figure}
\begin{center}
\epsfig{figure=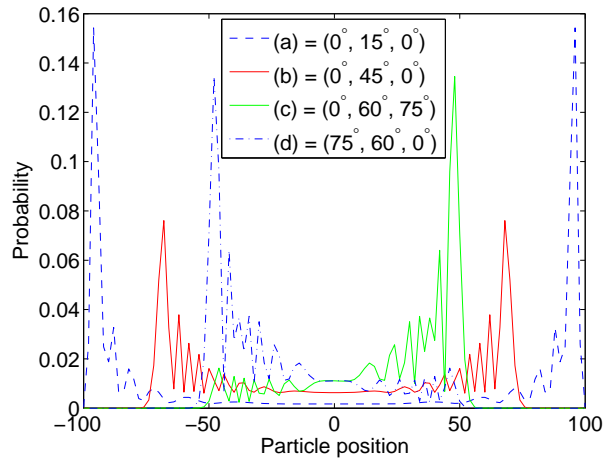, width=8.6cm}
\caption{\label{fig:qw}The spread of probability distribution for different value of $\theta$ using operator $U_{0, \theta, 0}$, is wider for (a) = $(0,  \frac{\pi}{12},  0)$ than compared to (b) = $(0,  \frac{\pi}{4}, 0)$.   Biasing the walk using $\zeta$ shifts the distribution to right, (c) = $(0, \frac{\pi}{3}, \frac{5 \pi}{12})$  and $\xi$ shifts to the left, (d) = $(\frac{5 \pi}{12}, \frac{\pi}{3}, 0 )$. The distribution is for 100 steps.}
\end{center}
\end{figure}

To  define  the  one-dimensional  discrete time  quantum (Hadamard) walk  we require the {\it  coin} Hilbert space $\mathcal H_{c}$  and the {\it position}  Hilbert space $\mathcal H_{p}$.  The $\mathcal  H_{c}$ is spanned by the internal 
(basis) state of the particle, $|0\rangle$ and $|1\rangle$, and  the 
$\mathcal  H_{p}$ is spanned  by the  basis state $|\psi_{i}\rangle$, 
$i \in \mathbb{Z}$. The total system is then in the space $\mathcal H
= \mathcal  H_{c} \otimes \mathcal  H_{p}$. To implement  the simplest
version of the quantum walk,  known as the Hadamard walk, the particle
at origin  in one of the basis state is evolved into the superposition of 
the $\mathcal H_{c}$ with equal probability, by applying the 
Hadamard operation, $H = \frac{1}{\sqrt 2} 
\left( \begin{array}{clcr}
 1  & &   1   \\
1  & &  -1 
\end{array} \right)$, such that, 

\begin{eqnarray}
\label{eq:shift}
(H\otimes \mathbbm{1})  (|0\rangle\otimes|\psi_{0}\rangle) =\frac{1}
{\sqrt 2}[|0\rangle+|1\rangle]\otimes|\psi_{0}\rangle \nonumber \\
(H\otimes \mathbbm{1}) (|1\rangle\otimes|\psi_{0}\rangle) =\frac{1}
{\sqrt 2}[|0\rangle-|1\rangle]\otimes|\psi_{0}\rangle.
\end{eqnarray}
The $H$ is then followed by the conditional shift operation $S$: 
conditioned on the internal state being $|0\rangle$ ($|1\rangle$) the particle moves to the left (right),
\begin{eqnarray}
\label{eq:condshift}
S = |0\rangle \langle 0|\otimes \sum_{i \in \mathbb{Z}}|\psi_{i-1}\rangle 
\langle \psi_{i} |+|1\rangle \langle 1 |\otimes \sum_{i \in \mathbb{Z}} 
|\psi_{i+1}\rangle \langle \psi_{i}| 
\end{eqnarray}
The operation $S$ evolves the particle into the superposition in position space. Therefore, each step of quantum  (Hadamard) walk is
composed of an application of $H$ and a subsequent $S$ operator to
spatially entangle $\mathcal H_{c}$  and $\mathcal H_{p}$. The process
of $W = S(H\otimes \mathbbm{1})$  is iterated without resorting to the
intermediate measurements  to realize a  large number of steps  of the
quantum walk. After the first  two steps of implementation of $W$, the
probability   distribution  starts  to   differ  from   the  classical
distribution. The probability  amplitude distribution arising from the
iterated  application  of  $W$  is significantly  different  from  the
distribution of the classical walk. The particle with initial coin state 
$|0\rangle$ ($|1\rangle$) drifts to the right (left). This asymmetry arises from the fact that the Hadamard operation treats the two states $|0\rangle$ and $|1\rangle$ differently, multiplies the phase by $-1$ only in case of state $|0\rangle$. To obtain left-right symmetry in the probability distribution, 
(b) in Fig. (\ref{fig:qw} b), one needs to start the walk with the partilce in the  symmetric superposition state of the coin, $|\Psi_{in}\rangle \equiv \frac{1}{\sqrt 2}[|0\rangle + i |1\rangle]\otimes |\psi_{0}\rangle$. 
\section{Generalized discrete time quantum walk}
\label{gen-qw}
The coin toss operation in  general can be  written as an arbitrary three  parameter $SU(2)$ operator  of the form, 
\be
U_{\xi,\theta,\zeta} \equiv \left( \begin{array}{clcr}  e^{i\xi}\cos(\theta)  & &   e^{i\zeta}\sin(\theta)   \\
e^{-i\zeta} \sin(\theta)  & &  -e^{-i\xi}\cos(\theta)
\end{array} \right),
\ee
\noindent the Hadamard operator,  $H = U_{0,  \frac{\pi}{4} ,0}$. By replacing  the Hadamard coin  with an operator $U_{\xi, \theta, \zeta}$,  we obtain the generalized quantum walk. For the analysis of the generalized quantum walk we consider the symmetric superposition state of the particle at the origin. By varying the parameter $\xi$ and $\zeta$ the results obtained for walker starting with one of the basis (or other nonsymmetric superposition) state can be reproduced. A  particle at origin in a symmetric superposition
state $|\Psi_{in}\rangle$, when subjected to a subsequent iteration of
$W_{\xi, \theta,  \zeta} = S(U_{\xi, \theta,  \zeta} \otimes {\mathbbm
1})$ implements a generalized discrete time quantum walk on a line. 
Consider an implementation of  $W_{\xi, \theta, \zeta}$, which evolves 
the walker to, 
\begin{eqnarray}
\label{eq:condshift2}
W_{\xi, \theta, \zeta}|\Psi_{in}\rangle =  \frac{1}{\sqrt 2}
[\left(e^{i\xi}  \cos(\theta)+ i e^{i\zeta} \sin(\theta)\right )
|0\rangle|\psi_{-1}\rangle \nonumber \\
+ \left( e^{-i\zeta}\sin (\theta) - i e^{-i\xi} 
\cos(\theta)\right) |1\rangle|\psi_{+1}\rangle ]. 
\end{eqnarray}
If $\xi=\zeta$, Eq. (\ref{eq:condshift2}) has left-right symmetry
in the position probability distribution,  but not
otherwise.  We thus find that  the generalized  $SU(2)$
operator as a quantum coin can bias a quantum walker in spite of the
symmetry of initial state of the particle. We return to this
point below.
\par
\begin{figure}
\begin{center}
\epsfig{figure=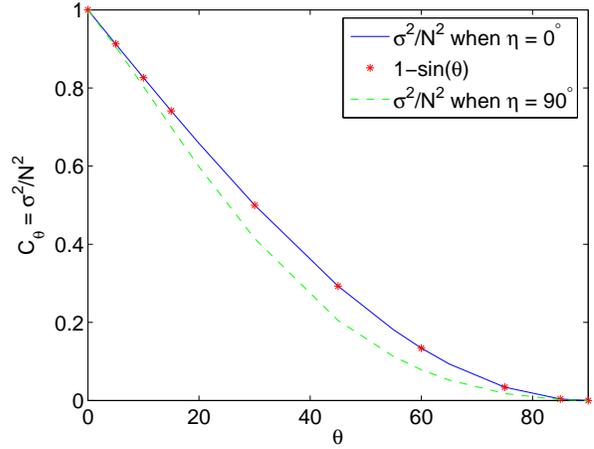,width=8.6cm}
\caption{\label{fig:CwithTheta}Variation of $C_{\theta}$ when $\eta = \lvert \xi - \zeta \rvert = 0^\circ$ from numerical integration to the function $(1-\sin(\theta))$ to which it fits. The effect of maximum biasing, $\eta = 90^\circ$ on $C_{\theta}$ is also shown and its effect is very small.}
\end{center}
\end{figure}

It is instructive to consider  the extreme values of the parameters in
the $U_{\xi, \theta, \zeta}$.  If $\xi=\theta=\zeta=0$, $U_{0, 0,0}=Z$, the Pauli $Z$ operation, then $W_{\xi, \theta, \zeta} \equiv
S$ and the two superposition states, $|0\rangle$ and
$|1\rangle$, move away from  each other without any diffusion and interference having high $\sigma^{2}= N^{2}$ .  On the other  hand, if  
$\theta= \frac{\pi}{2}$,  then  $U_{0, \frac{\pi}{2}, 0}=X$, the Pauli  $X$ operation, then the two  states cross each other
going  back and  forth, thereby  remaining  close to  position $i=0$ and  hence
giving very low $\sigma^{2} \approx 0$. These two extreme case are not
of  much  importance, but  they  define  the limits of the behavior.  Intermediate values of the $\theta$ between these extremes show intermediate drifts and quantum interference.
\begin{figure}
\begin{center}
\epsfig{figure=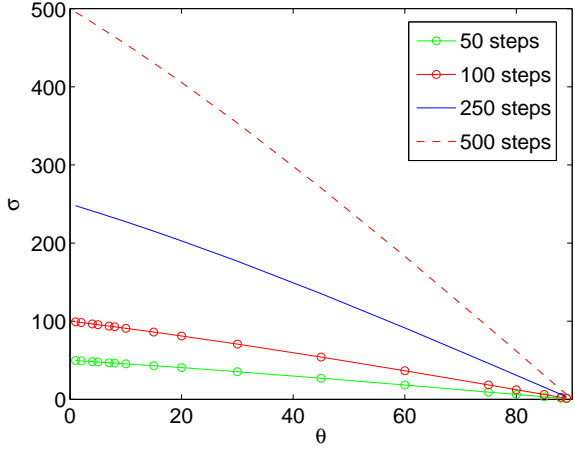, width=8.6cm}
\caption{\label{fig:qwsdtheta}A comparison of variation of $\sigma$  with  $\theta$ for different number of steps of walk using operator $U_{0,\theta, 0}$ using numerical integration.}
\end{center}
\end{figure}
In Fig.  (\ref{fig:qw}) we show the symmetric  distribution of quantum
walk at different values of $\theta$ by numerically evolving the density  
matrix. Fig. (\ref{fig:qwsdtheta}) shows the variation of  $\sigma$ with 
increase  in $\theta$ for quantum walk of different number steps with the operator, $U_{0,  \theta, 0}$. The change in the variance for  different value of the $\theta$ is attributed to the  change in the value of $C_{\theta}$, a constant for a given $\theta$, $\sigma^{2} = C_{\theta}N^{2}$, Fig. (\ref{fig:CwithTheta}). Therefore, starting from the Hadamard walk ($\theta=\frac{\pi}{4}; \xi=\zeta=0$), the variance can be increased ($\theta<\frac{\pi}{4}$) or decreased ($\theta > \frac{\pi}{4}$) respectively.
\par
In the analysis  of Hadamard walk on the line  in \cite{andris}, it is
shown  that after $N$ steps, the  probability distributed is spread over the interval $[\frac{-N}{\sqrt  2 }, \frac{N}{\sqrt 2 }]$   and shrink  quickly outside this  region. The  moments have been calculated  for asymptotically large number of steps $N$ and the variance is shown to vary as $\sigma^{2}(N)    =    \left(1-\frac{1}{\sqrt 2}\right)N^2$ \cite{andris}.
\par
\begin{figure}
\begin{center}
\epsfig{figure=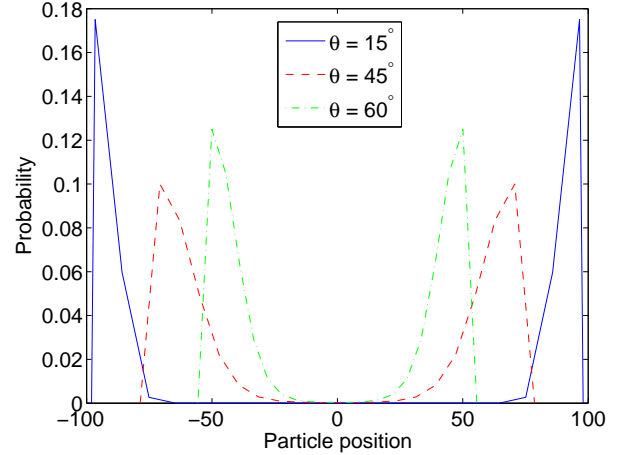, width=8.6cm}
\caption{\label{fig:qwfit}The probability distribution obtained using using Eq. (\ref{eq:proba1}) for different value of $\theta$. The distribution is for 100 steps.}
\end{center}
\end{figure}

The expression for the variance of the quantum  walk using $U_{0, \theta,  0}$ as a quantum  coin can be derived by using the approximate analytical function for the probability distribution $P(i)$ that fit the envelop of the quantum walk distribution obtained from the numerical integration technique for different values of $\theta$. For  a quantum  walk using $U_{0, \theta,  0}$ as  quantum  coin, after $N$ steps the probability distribution is spread over the interval $(-N\cos(\theta), N\cos(\theta))$ \cite{ashwin}. This is also verified by the analyzing the distribution obtained using the numerical integration technique. 
By assuming the value of the probability to be $0$ beyond $|N\cos(\theta)|$, the function that fits the probability distribution envelop is, 
\begin{eqnarray}
\int P(i) d i \approx \nonumber \\
\int_{-N\cos(\theta)}^{N\cos(\theta)} \frac{[1+\cos^{2}(2\theta)]e^{K(\theta)\left ( \frac{i^2}{N^2 \cos^2(\theta)} - 1 \right )}}{\sqrt N}  d i \approx 1,
\label{eq:proba1}
\end{eqnarray}
\noindent where, $K(\theta) =\frac{\sqrt N}{2}\cos(\theta)[1+\cos^{2}(2\theta)] [1 + \sin(\theta)]$ \cite{notes}. Fig. (\ref{fig:qwfit}) shows the probability distribution obtained by using the Eq. (\ref{eq:proba1}).
The interval $(-N\cos(\theta), N\cos(\theta))$ can be parametrised as a function of $\phi$, $i = f(\phi) = N \cos(\theta)\sin(\phi)$ where $\phi$ range from $-\frac{\pi}{2}$ to $\frac{\pi}{2}$. For a walk with coin $U_{0, \theta, 0}$, the mean of the distribution is zero and hence the variance can be analytically obtained by evaluating, 
\begin{eqnarray}
\sigma^{2} \approx \int_{-N\cos(\theta)}^{N\cos(\theta)} P(i) i^2 d i = \int_{-\frac{\pi}{2}}^{\frac{\pi}{2}} P(f(\phi))(f(\phi))^2 f^{\prime}(\phi) d\phi.
\label{eq:vari}
\end{eqnarray}
\begin{eqnarray}
\sigma^{2} \approx \int_{-\frac{\pi}{2}}^{\frac{\pi}{2}} \frac{(1+\cos^{2}(2\theta))}{\sqrt N}{e^{K(\theta)\left (\sin^2(\phi)- 1 \right )}}\left( N\cos(\theta)\sin(\phi)\right )^2 \nonumber \\
\times \left(N\cos(\theta)\cos(\phi) \right) d \phi = N^2 (1-\sin(\theta)).
\label{eq:vari1}
\end{eqnarray}
\begin{eqnarray}
\sigma^{2} = C_{\theta}N^{2} \approx (1-\sin(\theta)) N^2  .
\label{eq:vari2}
\end{eqnarray}
We also verify from the results obtained through numerical integration that $C_{\theta} = (1  - \sin(\theta))$, Fig. (\ref{fig:CwithTheta}).
\par
Setting $\xi \neq \zeta$ in $U_{\xi,\theta,\zeta}$ introduces asymmetry, biasing the walker. Positive $\zeta$  contributes  for  constructive  interference towards right and destructive interference  to the left, whereas vice versa for $\xi$. The inverse effect can be noticed when 
the $\xi$ and $\zeta$ are negative. As noted above, for $\xi = \zeta$, the evolution will again lead to the  symmetric probability distribution. Apart from a global phase, one can show that the coin operator 
\be
U_{\xi, \theta,  \zeta} \equiv U_{\xi - \zeta, \theta, 0} \equiv U_{0, \theta,  \zeta  - \xi}.
\ee  
In  Fig. (\ref{fig:qw}) we show the biasing  effect for 
$(\xi,  \theta, \zeta)  = (  0^\circ ,  60^\circ ,
75^\circ)$ and for $(75^\circ,  60^\circ , 0^\circ)$. The biasing does
not alter the width of the distribution in the position space but probability 
goes down as a function of $\cos(\eta)$ on one side and up as a function  
of $\sin(\eta)$ on the other side.  Where $\eta = \lvert \xi-\zeta \rvert$ . The mean value $\bar{i}$ of the distribution, which is zero for $U_{0, \theta, 0}$, attains some finite value with non-vanishing $\eta$, this contributes for an additional term in Eq. (\ref{eq:vari}),
\begin{eqnarray}
\sigma^{2} \approx \int_{-N\cos(\theta)}^{N\cos(\theta)} P(i) (i-\bar{i})^2 d i. 
\label{eq:varimean}
\end{eqnarray}
this contributes to a small decrease in the variance of the biased quantum walker, Fig. (\ref{fig:CwithTheta}).
\par
It  is  understood  that,  obtaining  symmetric  distribution  depends
largely on  the initial state of  the particle and this  has also been
discussed  in  \cite{ashwin,   andris,  tregenna,  konno}.  But  using
$U_{\xi, \theta,  \zeta}$ as coin operator, and
examining the walk evolution shows how non-vanishing $\xi$ and
$\zeta$ introduce bias. For example, the position probability
distribution in Eq. (\ref{eq:condshift2}) corresponding to
the left and right positions are $\frac{1}{2}[1 \pm \sin(2\theta)\sin(\xi - \zeta)]$,
which would be equal, and lead to a symmetric distribution, if and only 
if $\xi = \zeta$. The evolution of the state after $n$ steps,
$[W_{\xi, \theta,  \zeta}]^n |\Psi_{in}\rangle$ is
\begin{equation}
|\Psi(n)\rangle = \sum_{m=-n}^n 
(A_{m,n}|0\rangle|\psi_m\rangle + B_{m,n}|1\rangle|\psi_m\rangle)
\label{eq:lr}
\end{equation}
and proceeds according to the iterative relations,
\begin{subequations}
\label{eq:iter}
\begin{eqnarray}
A_{m,n} = e^{i\xi}\cos\theta A_{m-1,n-1} +
e^{i\zeta}\sin\theta B_{m-1,n-1} \\
B_{m,n} = e^{-i\zeta}\cos\theta A_{m-1,n-1} -
e^{-i\xi}\sin\theta B_{m-1,n-1}.
\end{eqnarray}
\end{subequations}
\par
A little algebra reveals that the solutions $A_{m,n}$ and
$B_{m,n}$ to Eqs. (\ref{eq:iter}) can be decoupled 
(after the initial step) and shown to satisfy
\begin{subequations}
\label{eq:iter0}
\begin{eqnarray}
A_{m,n+1}-A_{m,n-1} = 
\cos\theta(e^{i\zeta}A_{m-1,n} - e^{i\xi}A_{m+1,n}) \\
B_{m,n+1}-B_{m,n-1} = 
\cos\theta(e^{i\xi}B_{m-1,n} - e^{i\zeta}B_{m+1,n}).
\end{eqnarray}
\end{subequations}
For spatial symmetry from an initially symmetric superposition,
the walk should be invariant under an exchange of labels
$0 \leftrightarrow 1$, and hence should evolve $A_{m,n}$ and $B_{m,n}$ alike
(as in the Hadamard walk \cite{kni03}). From Eq. (\ref{eq:iter0}),
we see that this happens if and only if $\xi=\zeta$. 
\section{Entropy  of measurement}
\label{entropy}
As  an alternative measure  of position
fluctuation to variance, we consider the Shannon entropy of the
walker  position probability distribution  $p_i$ obtained  by tracing
over the coin basis: 
\be
H(i)= -\sum_i p_i \log p_i.
\ee
\begin{figure}
\begin{center}
\epsfig{figure=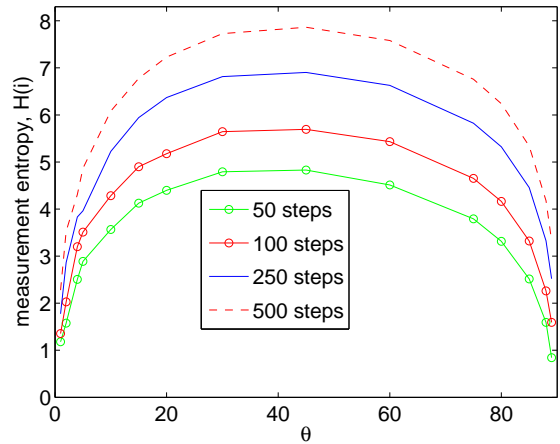, width=8.6cm}
\caption{\label{fig:qwEntropy}Variation of entropy of measurement $H(i)$ with $\theta$ for different number of steps. The decrease in $H(i)$ is not drastic till $\theta$ is close to $0$ or $\frac{\pi}{2}$.}
\end{center}
\end{figure}
The quantum walk with a Hadamard coin toss, $U_{0,  \frac{\pi}{4}, 0}$, has the maximum
uncertainty associated with the probability distribution and hence the
measurement  entropy  is  maximum.   For $\xi = \zeta =0$
and low  $\theta$, operator $U_{0,  \theta, 0}$ is almost  a Pauli $Z$
operation, leading to localization of walker at $\pm N$.   
At $\theta$ close to  $\frac{\pi}{2}$, with $\xi=\zeta=0$, 
$U$  approaches Pauli $X$  operation, leading to localization close to
the origin, and again, low entropy. However,
as $\theta$ approaches $\frac{\pi}{4}$, 
the splitting of amplitude in position space increases
towards the maximum. The  resulting enhanced diffusion is reflected in
the relatively large entropy at $\frac{\pi}{4}$, 
as seen in Fig.  (\ref{fig:qwEntropy}).  
Fig. (\ref{fig:qwEntropy})  is the measurement  entropy with variation
of  $\theta$ in the  coin $U_{0,  \theta, 0}$  for different  number of
steps of  quantum walk.  The decrease  in entropy from  the maximum by
changing $\theta$ on either side of $\frac{\pi}{4}$ is not drastic
untill the $\theta$  is close to $0$ or  $\frac{\pi}{2}$.  Therefore for
many practical purposes,  the small entropy can be  compensated for by
the relatively  large $C_{\theta}$, and hence  $\sigma^{2}$.  For many
other purposes, such as mixing of quantum walk on an $n$-cycle Cayley graph, it is ideal to adopt a  lower value of $\theta$.  The effect  of $\xi$ and $\zeta$ on  the  measurement  entropy   is  of  very  small  magnitude.  These parameters  do not  affect  the  spread of  the  distribution and  the variation in the height reduces the entropy by a very small fraction.
\section{Quantum walk on the  $n$-cycle and mixing time}
\label{cycle}
The $n$-cycle is
the simplest finite Cayley graph  with $n$ vertices.  This example has
most of the features of the walks on the general graphs.
\begin{figure}
\begin{center}
\epsfig{figure=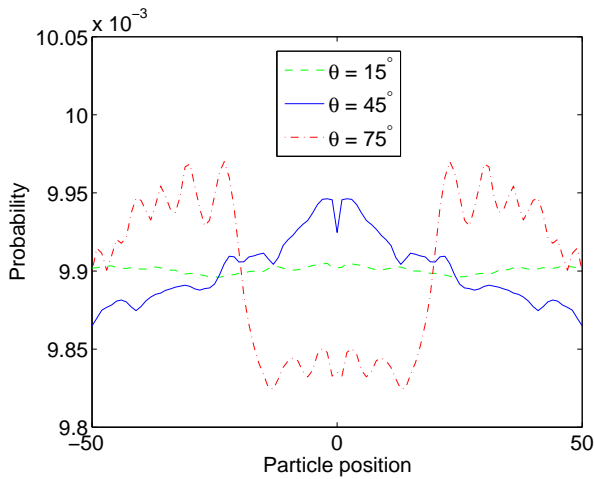, width=8.6cm}
\caption{\label{fig:mixing}A comparison of mixing time $M$ of the probability distribution of a quantum walker on a $n$-cycle for different value of $\theta$ using coin operation $U_{0,\theta,0}$, where $n$, the number of position, is $101$. Mixing is faster for lower value of $\theta$. The distribution is for 200 cycles.} 
\end{center}
\end{figure}
The  classical  random   walk  approaches  a  stationary  distribution
independent of  its initial  state on a  finite graph.   
Unitary (i.e., non-noisy) quantum walk,
does not converge to any stationary distribution.
But by defining a  time-averaged distribution, 
\be
\overline{P(i,T)} =
\frac{1}{T}  \sum_{t=0}^{T-1}  P(i,  t), 
\ee
\noindent obtained  by  uniformly picking a random time $t$ between $0$ and $(T-1)$, and evolving for $t$ time steps and  measuring to see which vertex it is at, a convergence in the  probability distribution can be  seen even in quantum  case. It has been shown  that the quantum walk on an $n$-cycle mixes in  time $M =
{\it O}(n \log n)$, quadratically faster than the classical case which
is ${\it  O}(n)$ \cite{dorit}.  From Eq. (\ref{eq:vari})  we know that
the quantum walk can be optimized for maximum variance and wide spread
in position  space, between $(-N \cos(\theta),  N \cos(\theta))$ after
$N$ steps. For a  walk on an $n$-cycle, choosing $\theta$  slightly above $0$
would give the maximum spread in the cycle during each cycle. Maximum spread during each cycle distributes the probability over the cycle faster and this
would optimize the mixing time. Thus optimizing mixing time with lower 
value of  $\theta$ can  in general  be applied  to most  of the  finite
graphs. For optimal mixing time, it turns out to be ideal to fix $\xi= \zeta$ in $U_{\xi, \theta,  \zeta}$, since  biasing  impairs a proper mixing.    Fig.  (\ref{fig:mixing})  is the  time  averaged probability distribution of a  quantum walk on an $n$-cycle graph  after $n \log n$ time where $n$ is $101$. It can be seen that the variation of the probability distribution over the position space is least for $\theta = 15^{\circ}$ compared to $\theta = 45^{\circ}$ and $\theta = 75^{\circ}$. 
\section{Quantum walk search}
\label{search}
A fast  and wide spread defines the effect of
the  search  algorithm. For the basic algorithm using discrete time quantum 
walk, two quantum coins are defined,  one  for a  marked  vertex and  
the other for an unmarked vertex. The three parameter of the $SU(2)$ 
quantum coin can be exploited for an optimal search.

\section{Conclusion}
\label{conclusion}
In  this paper we have  generalized the Hadamard walk to a general discrete 
time quantum walk with a $SU(2)$ coin. We conclude that the variance of quantum 
walk can  be optimized  by choosing  low $\theta$  without loosing  much on
measurement  entropy. The parameters  $\xi$  and  $\zeta$  introduce
asymmetry in  the position space probability distribution starting even from an initial symmetric superposition state. 
This asymmetry in the probability distribution is similar to the distribution obtained for a walk on a particle initially in a non-symmetric superposition state. Optimization of quantum search and mixing time on an $n$-cycle using  low $\theta$  is possible. The combination of  the parameters of the $SU(2)$ coin and the measurement entropy can be optimized to fit the physical system and for the relevant applications of the quantum walk on a general graph.
\bc
{\bf Acknowledgement}\\
\ec
CMC would like to thank the Mike and Ophelia Lazaridis fellowship for support. CMC and RL also acknowledge the support from CIFAR, NSERC, ARO/LPS grant 
W911NF-05-1-0469, and ARO/MITACS grant W911NF-05-1-0298.

\end{document}